%% file: main.tex
\documentclass[sigconf]{acmart} 
\acmConference[SEAMS 2022]{The 17th Symposium on Software Engineering for Adaptive and Self-Managing Systems}{May 21–29, 2022}{Pittsburgh, PA, USA}

\setcopyright{acmcopyright}
\copyrightyear{2022}
\acmYear{2022}

\acmConference[SEAMS 2022]{The 17th Symposium on Software Engineering for Adaptive and Self-Managing Systems}{May 21–29, 2022}{Pittsburgh, PA, USA}
\acmPrice{15.00}
\acmDOI{10.1145/3524844.3528081}
\acmISBN{978-1-4503-9305-8/22/05}

\usepackage{hyperref}
\usepackage{graphicx}
\usepackage{xcolor}
\usepackage{comment}
\usepackage{listings}
\usepackage{multicol}
\include{preamble}

\graphicspath{{figures/}}

\newcommand{\artifactName}{SEAByTE}
\definecolor{newC}{RGB}{250, 20, 20}
\newcommand{\new}[1]{\textcolor{black}{#1}}

\begin{document}

\title{\artifactName{}: A Self-adaptive Micro-service System Artifact for Automating A/B Testing}

\date{}

\author{Federico Quin}
\affiliation{%
    \institution{Katholieke Universiteit Leuven}
    \city{Leuven}
    \country{Belgium}
}
\email{federico.quin@kuleuven.be}

\author{Danny Weyns}
\affiliation{%
    \institution{Katholieke Universiteit Leuven, Belgium}
    \city{Linnaeus University}
    \country{Sweden}
}
\email{danny.weyns@kuleuven.be}

\begin{abstract}

Micro-services are a common architectural approach to software development today. An indispensable tool for evolving micro-service systems is A/B testing. In A/B testing, two variants, A and B, are applied in an experimental setting. By measuring the outcome of an evaluation criterion, developers can make evidence-based decisions to guide the evolution of their software. Recent studies highlight the need for enhancing the automation when such experiments are conducted in iterations. To that end, we contribute a novel artifact that aims at enhancing the automation of an  experimentation pipeline of a micro-service system relying on the principles of self-adaptation. Concretely, we propose \artifactName{}, an experimental framework for testing novel self-adaptation solutions to enhance the automation of continuous A/B testing of a micro-service based system. We illustrate the use of the \artifactName{} artifact with a concrete example. 

\end{abstract}

\maketitle

\section{Introduction}\label{sec:introduction}

Micro-services are nowadays a commonly used architectural approach to software development~\cite{PaoloDi2017}. A micro-service architecture  comprises small independent services that communicate over well-defined APIs. These services are usually owned by small, self-contained teams. %
Since each service performs a single function and runs independently, services can be easily updated, deployed, and scaled to meet changing demands. As such, a micro-service architecture naturally supports 
continuous deployment (CD)~\cite{978-3-319-08738}. CD is based on the principles of agile development~\cite{DINGSOYR20121213} and DevOps~\cite{MISHRA2020100308} and leverages on continuous integration (CI)~\cite{6802994} that automates tasks such as compiling code, running tests, and building deployment packages. Among the benefits of CI/CD are rapid innovation, shorter time-to-market, increased customer satisfaction, continuous feedback, and improved developer productivity. 

\new{One of the key concerns of CI/CD is increasing the agility of development teams in terms of testing, updating, maintaining and deploying software. Understanding user engagement and satisfaction of product features or variants plays an important role in this.}
A/B testing, also called bucket testing or controlled experimentation~\cite{Kohavi2007,Ros2018}, offers a solution to this concern: 
comparing two variants, A and B, and test if the statistical distribution of a property of A is different from that of B~\cite{Kohavi2017}.
By focusing on different properties in an experimental setting and systematically measuring the outcome, developers can
make evidence-based decisions to guide the software evolution. This is especially useful for (micro-)service based software with large user volumes~\cite{Ros2018}. Conducting experiments in iterations is known as continuous experimentation (CE)~\cite{FAGERHOLM2017292}. %
Recent studies highlight the need for enhanced automation of continuous experimentation as one of the key challenges in this area~\cite{RODRIGUEZ2017263,Ros2018}. 

In this paper, we contribute a novel artifact that aims at enhancing the automation of continuous experimentation of a micro-service system using the principles of self-adaptation. In particular, we propose \artifactName{}, an experimental framework that can be used for testing novel self-adaptation solutions to enhance the automation of continuous A/B testing of a micro-service based system. 

The remainder of this paper is structured as follows. In Section~\ref{sec:background}, we summarize the principles of micro-services and A/B testing and position \artifactName{} in the landscape of self-adaptive system artifacts.  Section~\ref{sec:artifact} presents the artifact with scenarios. Section~\ref{sec:usage} illustrates the application of the artifact for one of the scenarios. Section~\ref{sec:applicability} discusses the applicability of the artifact, Section~\ref{sec:future-research} discusses future research directions, and Section~\ref{sec:conclusions} wraps up.

\section{Background and Positioning of the Artifact}\label{sec:background}

We briefly introduce the basics of micro-services and A/B testing, and then highlight how \artifactName{} complements the existing artifacts for engineering self-adaptive systems.   

\subsection{Micro-services}\label{subsec:microservices}

A micro-service is a small independent piece of software that performs a single function (i.e., a business capability) and communicates with other micro-services via well-defined interfaces~\cite{Dragoni2017}. 
Each service in a micro-services architecture can be developed, deployed, operated, and scaled independently, without affecting 
other services. Services do not need to share any code with other services, hence they promote decentralized governance. When a change of a certain part of a micro-service based application is needed, only the related service(s) can be modified and redeployed without the need to modify and redeploy the entire application. 

Key players that migrated towards a micro-service architecture include Amazon, Netflix, eBay, and Twitter~\cite{8712375,Bogner21}. Yet, micro-service architecture also comes with challenges, such as determining the right size of micro-services, front-end integration, and failure management (need for self-healing)~\cite{8354433}. 

\subsection{A/B Testing}\label{subsec:abtesting}

A/B testing is a systematic approach to compare the use of two versions of a system and determining which of the two variants is preferred according to some criterion. More specifically, an A/B test consists of a randomized controlled experiment where experimental units (e.g., users) are assigned to one of two variants, called A and B, that are expected to influence some metric of interest. This metric, the overall evaluation criterion, provides a quantitative measure of the experiment‘s objective (in the form of a hypothesis). To compare the two variants, statistical hypothesis testing is commonly used. The choice of the tests (and its power) depends on the (assumed) distribution of the data. For instance, for Gaussian distributed data the unpaired t-test can be used. If the distribution is unknown, the Mann–Whitney U test can be used. 
A key aspect is the randomization applied to experimental units to map them to variants. Proper randomization is important to ensure that the populations assigned to the different
variants are statistically similar such that causal effects can be determined with
high probability. Prerequisites for A/B testing are: (1) experimental units can be assigned to different variants with no (or little) interference; (2) there are 
enough experimental units (e.g., users), usually thousands, (3) the key metrics are agreed upon and can be practically
evaluated, (4) experimental units can be easily assigned to variants (e.g., server-side software is much easier to change than client-side).
Continuous experimentation refers to pipelines of experiments. Often, the analysis results may trigger the need for additional (analysis) and follow-up experiments. Setting up such pipelines is a challenge for current experimentation platforms, see e.g.,~\cite{Gupta2019}. 

A/B testing has been extensively used, ranging from user interface element testing, testing product pricing, evaluating personalized recommendations, testing product features, and more generally evaluating the impact of changes made to software products and services. Online controlled experiments are today common practice and heavily used in practice~\cite{Gupta2019}.  

Multi-variant testing is similar to A/B testing, but the tests then use more than two versions at the same time. Yet, the current focus of \artifactName{} is on two-variant testing (A/B).

\subsection{Positioning of the Artifact}\label{subsec:positioning}

At the time of writing, the community of engineering self-adaptive systems has produced 27 artifacts. According to the SEAMS artifacts website\footnote{\url{https://www.hpi.uni-potsdam.de/giese/public/selfadapt/exemplars/}}\textsuperscript{,}\footnote{ZENODO: \url{https://zenodo.org/communities/seams/?page=1&size=20}} these artifacts provide ``examples, challenge problems, and solutions that the community can use to motivate research, exhibit and evaluate solutions and techniques, and compare results.'' 

Of the 27 artifacts, only three have a specific focus on service-based systems: Znn.com, TAS, and SWIM. 	
Znn.com~\cite{CMU-ISR-08-113} is centered on a webserver system that provides a simplified news site. The testing environment simulates the slash-dot effect which are periods of abnormally high traffic that overload the system. TAS~\cite{7194661} is an exemplar of a service-based system (SBS). SBSs are widely used in e-commerce, online banking, e-health and many other applications. In these systems, services offered by third-party providers are dynamically composed into workflows delivering complex functionality. SBSs increasingly rely on self-adaptation to cope with the uncertainties associated with third-party services, as the loose coupling of services makes online reconfiguration feasible. SWIM~\cite{8595390} is an exemplar for the evaluation and comparison of self-adaptation approaches for Web applications. The exemplar simulates a web application (simulating a 60-server cluster subject to millions of requests) that can be used as a target system with an external adaptation manager interacting with it through its TCP-based interface. 

While the existing artifacts produced by the community aim at supporting research on novel approaches for \textit{engineering self-adaptive systems}, \artifactName{} takes a different angle and provides an artifact that aims at supporting research on novel approaches for \textit{a key task of software engineers using self-adaptation}, in particular enhancing the automation of evolution with A/B testing. This aligns with recent initiatives of other researchers such as Self-Adaptation 2.0~\cite{9462033} that argues for an equal-to-equal relationship between self-adaptation and AI, benefiting one another. Another example is discussed in~\cite{Casimiro0GMKK21} where self-adaptation is applied to deal with degraded machine-learning components to maintain system utility. 

\section{\artifactName{}}\label{sec:artifact}

\begin{figure}[t!]
    \centering
    \includegraphics[width=\linewidth]{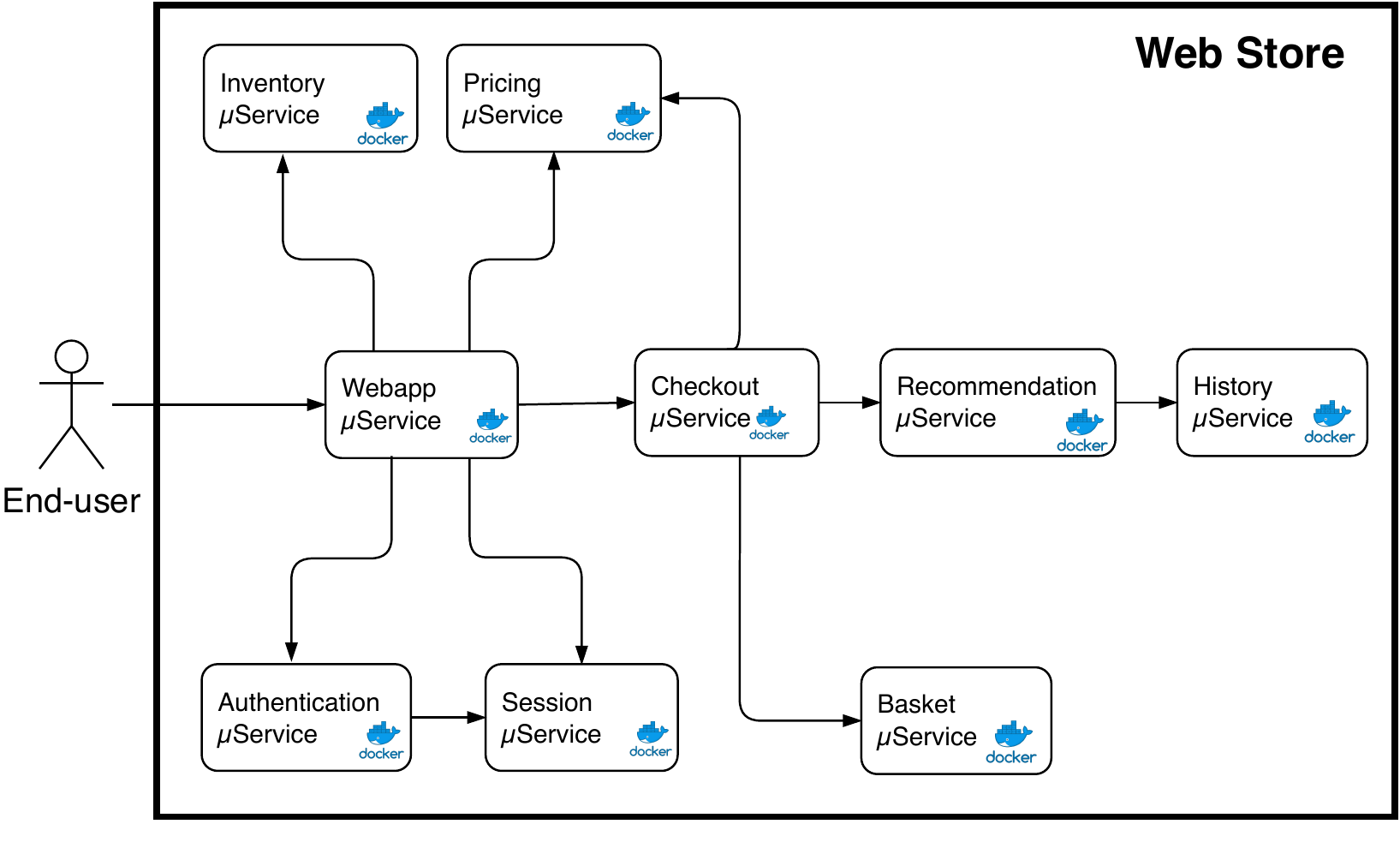}
    \caption{Architecture of the Web Store Application.}
    \label{fig:architecture-web-store}
\end{figure}

\artifactName{} provides an Internet Web Store composed of multiple micro-services.
Figure~\ref{fig:architecture-web-store} shows the architecture of the Web Store of \artifactName{}.  
The Web Store serves end-users that perform purchases via the  website of the store. When an end-user invokes a request, the Webapp service will authenticate the user, start a session, retrieve the price of the listed product, and update the inventory. Then the checkout service will be invoked that gives an overview of the products present in the basket of the user, provides recommendations for the user and finally adds the purchase to the history of the user. This closes the session. Practically, the web-services are implemented using the Java Spring framework, with each service running in a separate Docker container. The implementation of the Web Store artifact, an installation and usage guide, and a concrete example are available at the \artifactName{} website~\cite{artifact-website}.  

\subsection{Experimental Pipeline}

\artifactName{} aims to enhance the automation of the evolution of the micro-service system using a series of A/B tests. Central to this automation is an experimental pipeline that comprises two basic elements: experiments and transition rules. The artifact provides templates to define both experimental pipelines and transition rules. 

\subsubsection{Experiments}
An experiment comprises: 

\begin{itemize}
    \item \textit{ID}: identifier experiment 
    \item \textit{variantA}: the first variant
    \item \textit{variantB}: the second variant
    \item \textit{userProfile}: the profile specifying end-user behaviors  
    \item \textit{ABAssignment}: mapping of end-users to variants 
    \item \textit{samples}: the  number of samples used for the test
    \item \textit{metrics}: the set of metrics 
    \item \textit{statisticalTest}
    \begin{itemize}
    \item \textit{hypothesis}: the hypothesis of the experiment
    \item \textit{pValue}: the p-value for the test\footnote{\new{The specified p-value decides, after collecting the specified number of samples for both variants, if the test can be rejected based on the observed p-value. The result if the test will either be 'reject' or 'inconclusive'.}}
    \item \textit{type}: the type of statistical test of the experiment
    \item \textit{result}: the result variable of the experiment
    \end{itemize}
\end{itemize}

Listing~\ref{listing:experiment} shows an example of an experiment with ID ``Upgrade v1.0.0 - v1.1.0'' for the artifact. 
\vspace{5pt}
\begin{lstlisting}[language=JSON, caption={Example specification of  experiment in \artifactName{}.}, frame=single, captionpos=b, label={listing:experiment}]
{
  "Upgrade v1.0.0 - v1.1.0": {
    "variantA": "ws-recommendation-service:1.0.0",
    "variantB": "ws-recommendation-service:1.1.0",
    "userProfile": "Standard",
    "ABAssignment": {
      "weightA": 50,
      "weightB": 50
    },
    "samples": 20000,
    "metrics": ["ResponseTime_A", "ResponseTime_B"],
    "statisticalTest": {
        "hypothesis": "ResponseTime_A == ResponseTime_B",
        "pValue": 0.025,
        "type": "welsh's t-test",
        "resultingVariable": "result-wt-test"
    }
  }
}
\end{lstlisting}

This experiment compares the performance of two variants of the Web Store that implement different versions of the recommendation service, see Figure~\ref{fig:architecture-web-store-ab}. Version 1.0.0 of the recommendation service uses the history to provide recommendations to the user, while version 1.1.0 exploits also information of the current purchase to provide recommendations. The experiment uses a user-profile that generates user requests. The requests are randomly assigned to the two versions, 50\% each. The experiment tracks the requests with their response-time. The statistical test compares the mean values of the response times of the invocations of both versions using a Welsh's t-test~\cite{34.1-2.28}. The artifact uses the Apache Commons Mathematics Library for statistical testing~\cite{Commons-Math}. The result of the test (accept or reject the hypothesis) is recorded in the resulting variable.

\subsubsection{Transition Rules}
A transition rule comprises: 

\begin{itemize}
    \item \textit{ID}: identifier of the rule
    \item \textit{fromExperiment}: identifier current experiment 
    \item \textit{toExperiment}: identifier of the next experiment  
    \item \textit{conditions}: the conditions to make a transition from the current experiment to the next experiment (that depends on the result of the current experiment). 
\end{itemize}

\textit{to-experiment}\,=\,\textit{"end}" is a reserved value that indicates the end of an experimental pipeline and returns the control to the operator. Empty \textit{conditions} indicate that the transition from the current to the next experiment is taken unconditionally.

Listing~\ref{listing:rule} shows an example of the specification of a transition rule for the artifact. 
\vspace{5pt}
\begin{lstlisting}[language=JSON, caption={Example specification of a transition rule.}, frame=single, captionpos=b, label={listing:rule}]
{
  "Performance OK" : {
    "fromExperiment": "Upgrade v1.0.0 - v1.1.0",
    "toExperiment": "Clicks v1.0.0 - v1.1.0",
    "conditions": [{
       "leftOperand": "result-wt-test",
       "operator": "!=",
       "rightOperand": "reject"
    }]
  }
}
\end{lstlisting}

This transition rule is applied once the Upgrade v1.0.0 - v1.1.0  experiment is finished. It checks whether the result of the test was accepted, and if so, it starts the Clicks v1.0.0 - v1.1.0 experiment. 

\subsubsection{Experimental Pipeline}
Finally, an experimental pipeline comprises the following: 

\begin{itemize}
    \item setup: identifier of the setup for the pipeline\footnote{A setup comprises the necessary steps to be taken to get the managed system ready for conducting experiments (further explained in Section~\ref{sec:abcomponent-setup}).}
    \item start: the identifier of the experiment that starts the pipeline
    \item experiments: the set of experiment identifiers
    \item rules: a set of transition rule identifiers
\end{itemize}

\begin{figure}[h!]
    \centering
    \includegraphics[width=\linewidth]{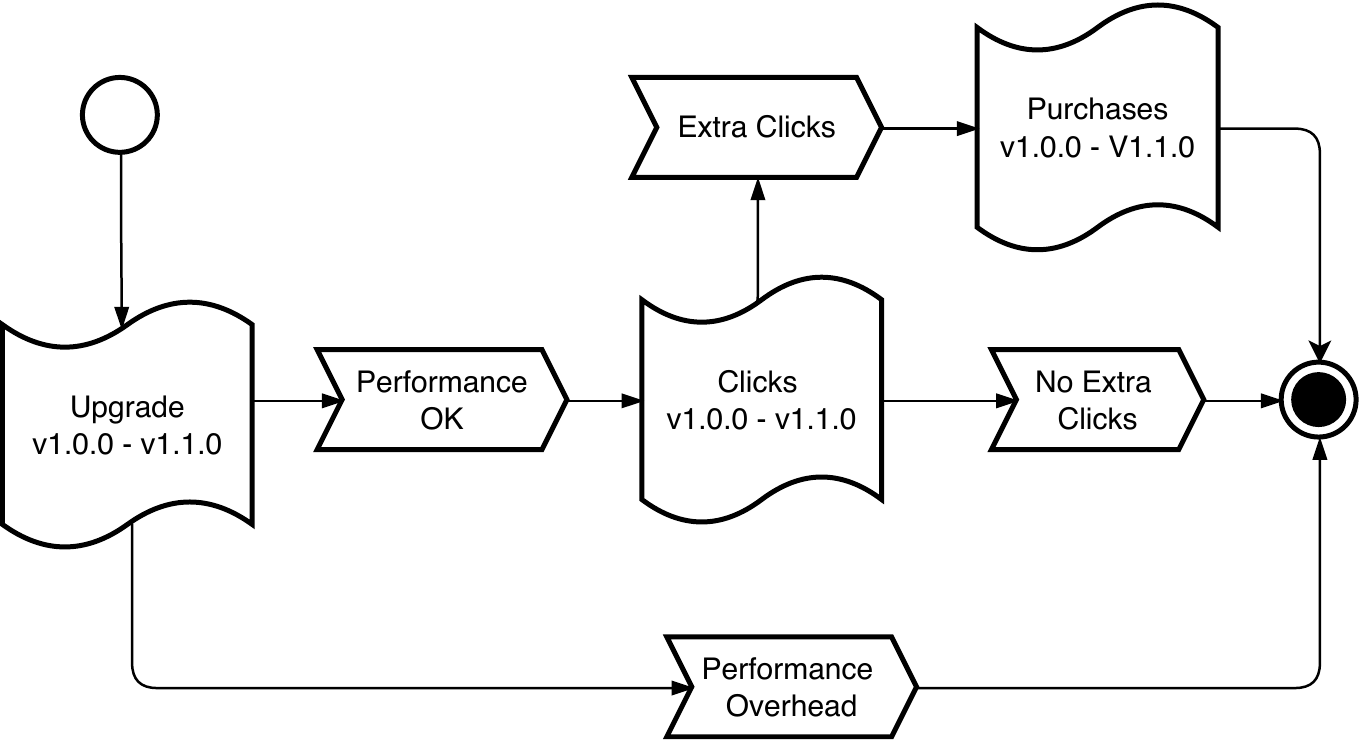}
    \caption{Example of experimental pipeline.}
    \label{fig:pipeline}
\end{figure}

Figure~\ref{fig:pipeline} shows a graphical representation of an example of an experimental pipeline, and Listing~\ref{listing:pipeline} shows how this pipeline is specified for the artifact. 
 
The experiment starts with preparing the Web Store according to the specified setup. Then, the Upgrade v1.0.0 - v1.1.0 experiment that checks the performance overhead in response time that is generated by the upgrade of the Web Store to version v1.1.0 is executed. If the performance is acceptable (i.e.,  version 1.1.0 of the Web Store generates no significant overhead in response time compared to version 1.0.0), the transition via the rule ``Performance OK'' is taken and the pipeline starts the Clicks v1.0.0 - v1.1.0 experiment. Otherwise, the transition via the rule ``Performance Overhead'' is taken and the execution of the pipeline ends. The Clicks experiment checks whether users perform a significant number of extra clicks on recommendations for version 1.1.0 compared to version 1.0.0. If this is not the case, the transition 
``No Extra Clicks'' is taken and the execution of the pipeline ends. Otherwise, the Purchases v1.0.0 - v1.1.0 experiment is started via the ``Extra Clicks'' rule. This last experiment compares the purchase behavior of the user with both variants and reports the result to the operator, completing the execution of the experimental pipeline. Based on the results the stakeholders can then decide whether or not to upgrade the Web Store, or set up additional experiments if needed. 
\vspace{5pt}
\begin{lstlisting}[language=JSON, caption={Example specification of experimental pipeline.}, frame=single, captionpos=b, label={listing:pipeline}]
{
  "setup": "Recommendation_upgrade",
  "start": "Upgrade v1.0.0 - v1.1.0",
  "experiments": [
    "Upgrade v1.0.0 - v1.1.0",
    "Clicks v1.0.0 - v1.1.0",
    "Purchases v1.0.0 - v1.1.0"
  ],
  "rules": [
    "Performance OK", 
    "Performance Overhead", 
    "Extra Clicks", 
    "No Extra Clicks"
  ]
}
\end{lstlisting}

\subsection{Architecture \artifactName{}}

\begin{figure*}
    \centering
    \includegraphics[width=0.85\linewidth]{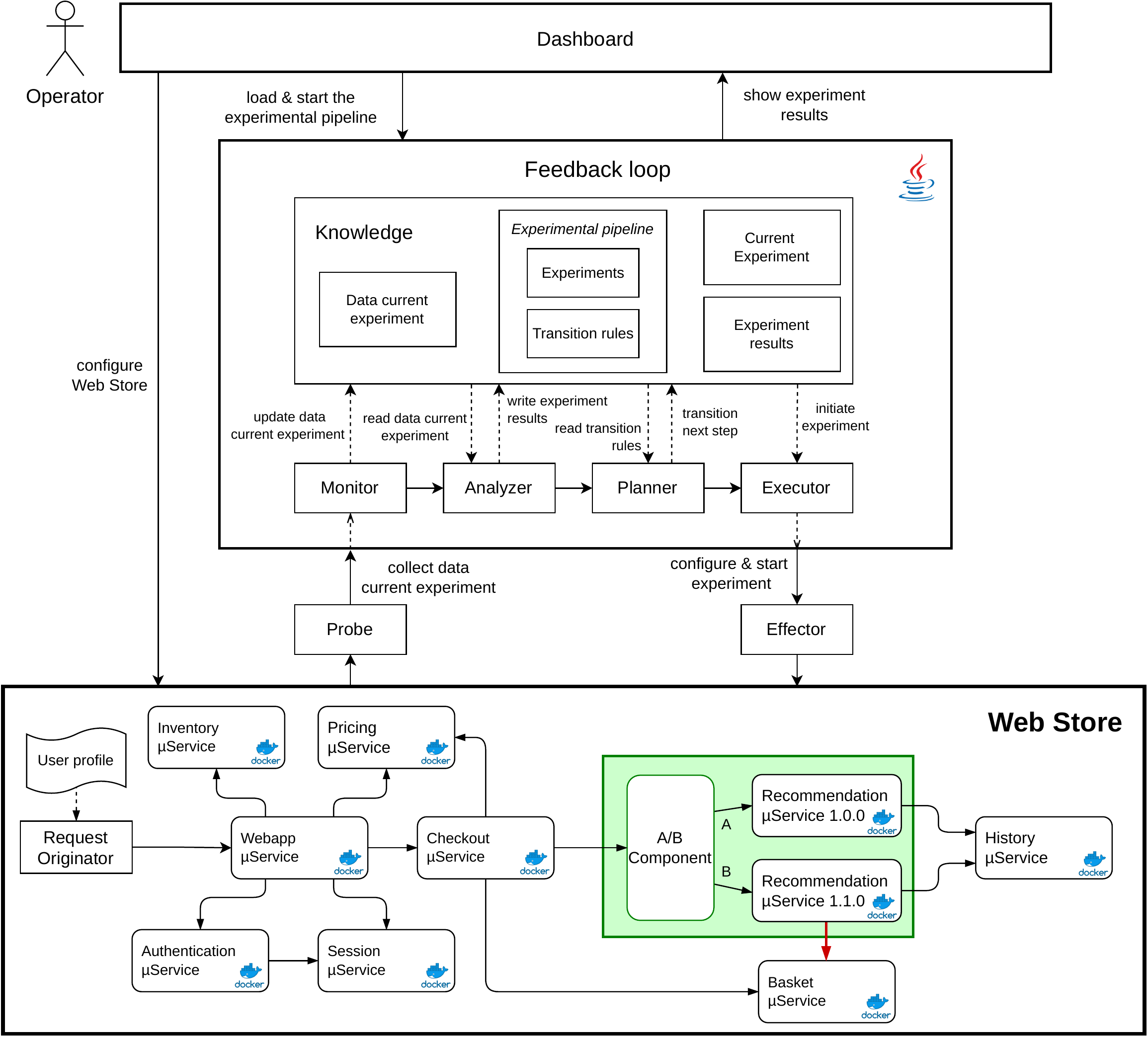}
    \caption{Architecture of \artifactName{}.}
    \label{fig:architecture-web-store-ab}
\end{figure*}

\artifactName{} adds a MAPE-K based feedback loop~\cite{Kephart,empiricalMAPE} on top of the Web Store application that supports the automatic execution of an experimental pipeline. 
Figure~\ref{fig:architecture-web-store-ab} shows the architecture of \artifactName{} illustrated for a scenario with two variants of the Recommendation service, A and B. We explain the main components now.

\subsubsection{User Profile} A user profile represents end-users of the Web Store; it defines the behavior of the users. Behavior refers to the actions taken by users and includes both the invocation of requests and feedback provided by users based on the completed requests. \artifactName{} provides a template to specify user profiles. Listing~\ref{listing:user-profile} shows an example instance of the profile template.

The example shows the profile for a standard user that specifies different aspects of the behavior of the users when visiting the website of the store: purchasing goods, browsing the store, exploring recommendations, and their effect. A profile can include different types of users. User profiles can be derived from different sources such as domain knowledge, historical data, or based on tests. \new{The request originator component exploits the user profile to generate the requests for the web store according to the specified parameters in the profile.}
 \vspace{5pt}
\begin{lstlisting}[language=JSON,
caption={Example specification of a user profile.},
frame=single,
captionpos=b,
label={listing:user-profile}]
{
  "user-profile-regular" : {
    "Standard": {
      "count": 1000,
      "mean-seconds-between-request": 15,
      "probability-purchase": 0.25,
      "recommendation-click-probability": 0.2,
      "recommendation-purchase-probability": 0.05,
      "bonus-recommendation-click-B": 0.1,
      "bonus-recommendation-purchase-B": 0.05
    },
    "frivolous": { 
      ...
    }
  }
}
\end{lstlisting}

\subsubsection{A/B Component}\label{sec:abcomponent-setup} The A/B component determines the test setup with A and B variants and manages the routing of invocations to the two variants. The A/B test setting will be configured before the execution of the experimental pipeline starts (via the dashboard, see below). The concrete routing to the A and B variants will be configured for each concrete experiment (the responsibility of the effector, see below).

\subsubsection{Probe and Effector} The probe is responsible for collecting data of experiments, while the effector is responsible for setting up the web store and managing the AB component. Listing~\ref{listing:probe-effector} shows a basic API of the probe and effector.
 \vspace{5pt}
\begin{lstlisting}[caption={The provided API for the probe and effector.}, frame=single, captionpos=b, label={listing:probe-effector}, basicstyle=\normalfont\ttfamily\footnotesize, breaklines=true, morekeywords={List, URLRequest, String, void, int}, keywordstyle={\color{eclipseStrings}},
emph={Probe, Effector},
emphstyle={\bfseries}]
Probe
  + List<URLRequest> getRequestHistory(String ABName, String variant)
  
Effector
  + void clearABComponentHistory(String ABName)
  + void setABRouting(String ABName, int a, int b)
  + void deploySetup(String setupName)
  + void removeSetup(String setupName)
\end{lstlisting}

The probe provides a method to collect the history of the requests of the active experiment of the service invocations per variant. A request contains information about the response time, the requested URL and the client ID. 
The effector enables setting up an experiment, including configuring the A and B setting and the routing for the variants, and clearing the history of the AB component.  

\subsubsection{Feedback Loop} The feedback loop is responsible for executing an experimental pipeline, see Figure~\ref{fig:architecture-web-store-ab}. The knowledge comprises a specification of the 
pipeline with the experiments and transition rules, the configuration of the current experiment, a repository to store data of the current experiment, and the experiment results.
The monitor collects the data of the running experiment and updates the knowledge repository accordingly. When the experiment is completed, the analyser applies the statistical test and writes the experiment result to the knowledge. The planner then applies the transition rules to the result of the last experiment and determines the next step in the experimental pipeline. Based on that, the executor initiates the next experiment, or alternatively, the experimental pipeline ends. The artifact provides abstract implementations of the feedback loop elements with an example. 

\begin{table*}[t!]
\caption{Test scenarios for \artifactName{}}\label{tab:scenarios}\vspace{-7pt}
\begin{center}
\begin{tabular}{p{0.3cm}p{3cm}p{8.5cm}p{4.6cm}}
\hline\noalign{\smallskip}
ID & Name & Description & Metric\\
\noalign{\smallskip}\hline\noalign{\smallskip}
S1 & Basic service upgrade & Experimental pipeline with two versions of a single micro-service & Response time of service invocations \\
S2 & Advanced service\newline upgrade & Experimental pipeline with two versions of a single micro-service & Response time of service invocations and user behavior\\
S3 & Setting product price & Experimental pipeline that determines the right price for a product with the aim to maximize the revenue & Appreciation of the price of\newline purchases by the users \\
S4 & Basic segmentation  & 
Experimental pipeline that determines  a segmentation strategy for two variants based on the age of end-users & Preferences of variants using user feedback\\
S5 & Advanced\newline segmentation  & 
Experimental pipeline that determines  a segmentation strategy for two variants based on the age and geographic location of end-users & Preferences of variants using user feedback   
\\
\noalign{\smallskip}\hline
\end{tabular}
\end{center}
\end{table*}

\begin{figure}[h!]
    \centering
    \includegraphics[width=.8\linewidth]{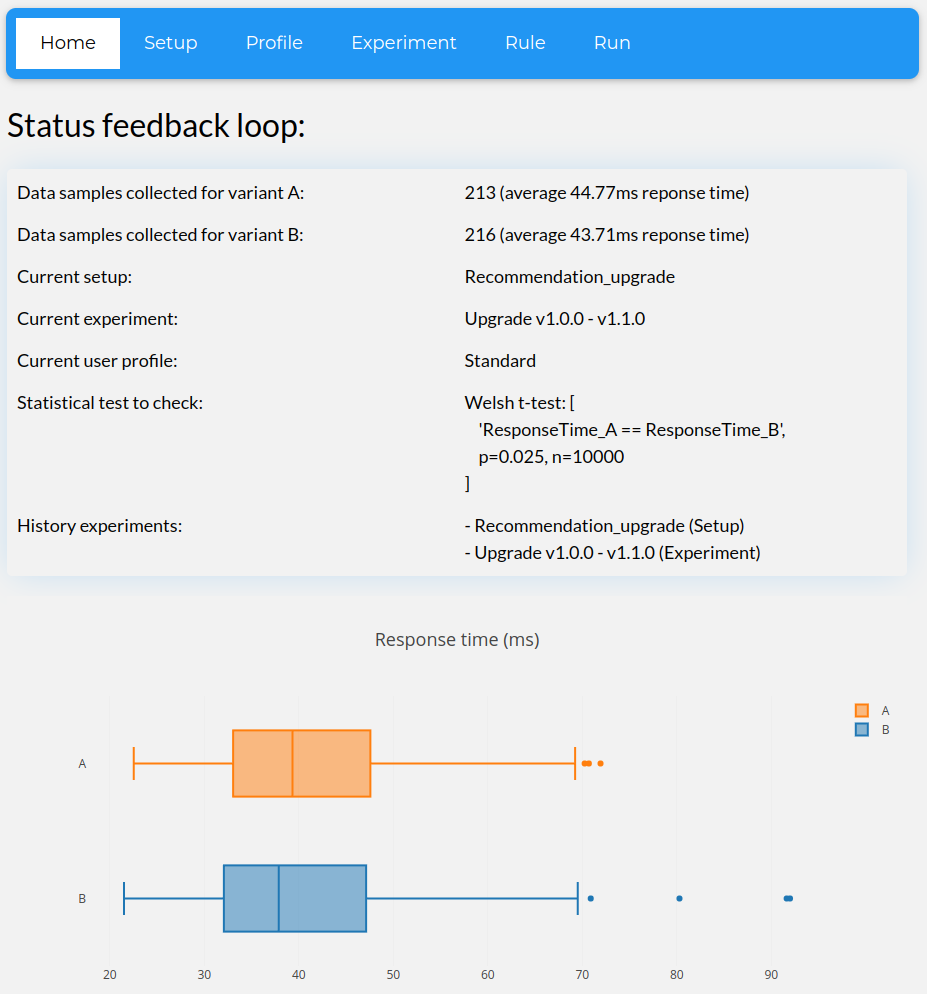}
    \caption{Excerpt of the dashboard of \artifactName{}.}
    \label{fig:dashboard-excerpt}
\end{figure}

\subsubsection{Dashboard}

The dashboard gives the operator access to the test environment, enabling: (1) configuration of the Web Store to run an experimental pipeline, (2) loading and starting an experimental pipeline, (3) monitoring the execution of an experimental pipeline, and (4) showing the 
results. Figure~\ref{fig:dashboard-excerpt} shows an excerpt of the dashboard 
to monitor the progress of the execution of an experimental pipeline. The box plot at the bottom 
shows a live representation of the response times for versions A and B.

\subsection{Test Scenarios} Table~\ref{tab:scenarios} shows the test scenarios supported by \artifactName{}. Each scenario introduces increasingly difficult challenges. The challenges focus on three key aspects of A/B testing: system upgrades (S1 and S2), price setting (S3), and segmentation (S4 and S5). The artifact provides a full specification for concrete instances of scenarios S1 and S2, and a basic implementation supporting the other scenarios.

\section{Experimentation with the Artifact}\label{sec:usage}

\subsection{Workflow to use the artifact}

Using the artifact comprises the following steps: 

\begin{itemize}
    \item Specify the experimental pipeline with the experiments and transition rules 
    \item Configure the Web Store with the variants to be tested 
    \item Configure the feedback loop to prepare the execution of the experimental pipeline 
    \item Load the experimental pipeline via the dashboard  
    \item Deploy the Web Store configuration and the feedback loop via the dashboard\new{, and start the experimental pipeline}
    \item Collect the experiment results and make a decision 
\end{itemize}

\subsection{Results}

We applied the steps above for the configuration with two variants of the recommendation service shown in Figure~\ref{fig:architecture-web-store-ab}. We applied the experimental pipeline shown in Figure~\ref{fig:pipeline} using the specifications of the different elements illustrated above.

We conducted the ``Upgrade v1.0.0 - v1.1.0'' experiment that compared the distributions of the response time of the two variants for 20k samples. 
After the 20k samples were collected, we observed that the null hypothesis (response time of v1.1.0 similar to response time v1.0.0) could not be rejected. The pipeline thus continues with the next experiment: ``Clicks v1.0.0 - v1.1.0''. For more details and additional results, we refer to the project website~\cite{artifact-website}.

\section{On the Applicability of \artifactName{}}\label{sec:applicability} 

\artifactName{} targets the automation of continuous experimentation of micro-service systems. The artifact aims for a high level of reality of a practical micro-services system, yet, several aspects of real world systems are emulated, including user behavior and system load. The artifact supports A/B testing for both functional and non-functional aspects. It also supports human involvement in the form of human feedback, which is a crucial aspect of A/B testing in practice. 

\artifactName{} comes with five scenarios; the code and specifications of two scenarios is shipped with the artifact. For the other scenarios, the artifact provides a basic implementation. Yet, the artifact does not exclude research on new scenarios. 

It is important to balance the design of an experimental pipeline. Automatically running multiple experiments may result in negative experience for some users~\cite{Kohavi2013}. Furthermore, it may accumulate errors of statistical tests, which needs to be considered when defining p-values for subsequent tests.

\section{Future research directions}\label{sec:future-research}

The current version of \artifactName{} supports the basics for setting up experimental pipelines for A/B testing using a feedback loop. We highlight several opportunities for future research beyond these basics. 
A first opportunity is to incorporate user feedback into the experimental pipeline, akin to one of the test scenarios presented in Table~\ref{tab:scenarios}. 
A second opportunity is to identify classes of users to target A/B tests for specific subgroups. Here self-adaptation can be combined with machine learning to identify such classes and determine which class users belong to~\cite{quin2019efficient,gheibi2021}. Incorporating user feedback in this setting can also play a crucial role in developing a suitable adaptation strategy.
A third opportunity for future research could be the automatic identification of A/B experiments that can be setup and run in the application. This can range from trying out different GUI layouts to testing different algorithms (similar to the recommendation algorithm tested in Section~\ref{sec:artifact}). A fourth and last opportunity is supporting multiple experiments simultaneously. Since running multiple A/B experiments can be a risk (experiments may affect each other), care has to be taken. Self-adaptation is a perfect candidate to deal with such concerns. Self-adaptation can also play a crucial role for dealing with a large space of potential experiments that need to run. Self-adaptation can further help in identifying the experiments that are most likely to have a high impact on the overall business goals of the system, and determining in which order to run the experiments.

\section{Conclusions}\label{sec:conclusions}

We presented \artifactName{}, a novel artifact that applies self-adaptation~\cite{weyns2020book} to enhance the automation of A/B testing to support the evolution of micro-service systems. In contrast to existing artifacts that target novel approaches to engineer self-adaptive systems, \artifactName{} exploits self-adaptation as a means to solve a key task of software engineers of service-based systems: automating continuous experimentation. 
We hope that the research community will use the \artifactName{} to evaluate research advances in the application of self-adaptation to support software engineers of micro-service systems.
The artifact is available via the project website~\cite{artifact-website}.

\onecolumn

\begin{multicols}{2}

\bibliographystyle{ieeetr}
\bibliography{main}

\end{multicols}
\end{document}

%% file: preamble.tex
\definecolor{reviewColor}{RGB}{220, 10, 10}

\definecolor{deepred}{rgb}{0.6,0,0}
\definecolor{eclipseStrings}{RGB}{42,0.0,255}
\definecolor{eclipseKeywords}{RGB}{127,0,85}
\colorlet{numb}{magenta!60!black}

\lstdefinelanguage{JSON}{
	basicstyle=\normalfont\ttfamily\footnotesize,
	commentstyle=\color{gray}, 
	stringstyle=\color{eclipseStrings},
	showstringspaces=false,
	breaklines=true,
	string=[s]{"}{"},
	literate=
	*{0}{{{\color{deepred}0}}}{1}
	{1}{{{\color{deepred}1}}}{1}
	{2}{{{\color{deepred}2}}}{1}
	{3}{{{\color{deepred}3}}}{1}
	{4}{{{\color{deepred}4}}}{1}
	{5}{{{\color{deepred}5}}}{1}
	{6}{{{\color{deepred}6}}}{1}
	{7}{{{\color{deepred}7}}}{1}
	{8}{{{\color{deepred}8}}}{1}
	{9}{{{\color{deepred}9}}}{1}
}